\documentclass[12pt,preprint]{aastex}

\def\feh{{\rm[Fe/H]}}
\def\alphafe{{\rm[\alpha/Fe]}}
\def\ltsima{$\; \buildrel < \over \sim \;$}
\def\simlt{\lower.5ex\hbox{\ltsima}}
\def\gtsima{$\; \buildrel > \over \sim \;$}
\def\simgt{\lower.5ex\hbox{\gtsima}}

\shorttitle{Chemical Abundances in the Sgr dSph}
\shortauthors{Smecker-Hane \& McWilliam}

\begin{document}

\newcommand{\znh}{[{\rm Zn/H}]}
\newcommand{\msol}{M_\odot}
\newcommand{\etal}{et al.\ }
\newcommand{\delv}{\Delta v}
\newcommand{\kms}{km~s$^{-1}$ }
\newcommand{\cm}[1]{\, {\rm cm^{#1}}}
\newcommand{\N}[1]{{N({\rm #1})}}
\newcommand{\e}[1]{{\epsilon({\rm #1})}}
\newcommand{\f}[1]{{f_{\rm #1}}}
\newcommand{\rAA}{{\AA \enskip}}
\newcommand{\sci}[1]{{\rm \; \times \; 10^{#1}}}
\newcommand{\ltk}{\left [ \,}
\newcommand{\ltp}{\left ( \,}
\newcommand{\ltb}{\left \{ \,}
\newcommand{\rtk}{\, \right  ] }
\newcommand{\rtp}{\, \right  ) }
\newcommand{\rtb}{\, \right \} }
\newcommand{\ohf}{{1 \over 2}}
\newcommand{\nohf}{{-1 \over 2}}
\newcommand{\rhf}{{3 \over 2}}
\newcommand{\smm}{\sum\limits}
\newcommand{\perd}{\;\;\; .}
\newcommand{\cmma}{\;\;\; ,}
\newcommand{\intl}{\int\limits}
\newcommand{\mkms}{{\rm \; km\;s^{-1}}}
\newcommand{\ew}{W_\lambda}


\title{The Complex Chemical Abundances and Evolution of the Sagittarius Dwarf 
Spheroidal Galaxy\altaffilmark{1}}

\author{Tammy A. Smecker-Hane\altaffilmark{2} and Andrew McWilliam\altaffilmark{3}}

\altaffiltext{1}{Data presented herein were obtained at the W.M.~Keck 
Observatory, which is operated as a scientific partnership among the 
California Institute of Technology, the University of California and 
the National Aeronautics and Space Administration. The Observatory was 
made possible by the generous financial support of the W.M.~Keck Foundation.}

\altaffiltext{2}{Department of Physics \& Astronomy, 4129 Frederick Reines Hall,
University of California, Irvine, CA 92697--4575; {\it tsmecker@uci.edu}}

\altaffiltext{3}{The Observatories of the Carnegie Institute of Washington,
813 Santa Barbara St., Pasadena, CA 91101--1292; {\it andy@ociw.edu}}

\begin{abstract}

We report on the chemical abundances derived from high-dispersion spectra
of 14 red giant stars in the Sagittarius dwarf spheroidal (Sgr dSph) galaxy.
The stars span a wide range of metallicities, $-1.6 \leq \feh \leq -0.1$ dex, and 
exhibit very unusual abundance variations. 
For metal-poor stars with $\feh < -1$, $\alphafe \approx +0.3$ 
similar to Galactic halo stars, but for more metal-rich stars the 
relationship of [$\alpha$/Fe] as a function of [Fe/H] is {\it lower} 
than that of the Galactic disk by 0.1 dex. The light elements 
[Al/Fe] and [Na/Fe] are sub-solar by an even larger amount, approximately $0.4$ dex.
The pattern of neutron-capture heavy elements, as indicated by [La/Fe] and
[La/Eu], shows an increasing $s$-process component with increasing [Fe/H], 
up to [La/Fe] $\sim +0.7$ dex for the most metal-rich Sgr dSph stars.  
The large [La/Y] ratios show that the $s$-process enrichments came from
the metal-poor population.  We can best understand the observed abundances 
with a model in which the Sgr dSph formed stars over a many Gyr 
and lost a significant fraction of its gas during its evolution.
Low-mass, metal-poor, AGB stars polluted the more metal-rich stars with 
$s$-process elements, and type~Ia SN from  low-mass progenitors 
enriched the ISM with iron-peak metals.  The type~II/type~Ia SN ratio 
was smaller than in the Galactic disk, presumably due to a slower star
formation rate; this resulted in the observed low [$\alpha$/Fe], 
[Al/Fe] and [Na/Fe] ratios. The fact that Sgr stars span such a wide range
in metallicity leads us to conclude that their age spread is even larger
than previously inferred. We derive ages for these red giants using the
Padova models (Girardi et al.~2000). The ages span $\sim 0.5$ to 13
Gyr, which implies a very long duration of star formation in the central 
regions of the Sgr dSph.

\end{abstract}

\keywords{stars: abundances --- galaxies: abundances --- galaxies: evolution ---
galaxies: dwarf --- galaxies: individual (Sgr dSph)}

\section{Introduction}

The Sagittarius dwarf spheroidal galaxy (Sgr dSph) is
currently being ripped apart and accreted onto the Milky Way (Ibata et al. 1997,
and references therein).  Tidal debris from the Sgr dSph appears to litter the
Galaxy tracing out Sgr's orbit (Dohm-Palmer et al.~2000, Newberg et al. 2002).
How many and how frequently have mergers
of dSphs effected the evolution of the Galaxy?  When did they occur? We can
answer these questions by determining the distributions of
ages and chemical abundances in dSphs and comparing them with 
those of Galactic stars (Unavane, Wyse \& Gilmore 1996, Shetrone, C\^{o}t\'e, \&
Sargent 2001, Fulbright 2002).
Abundance ratios such as [$\alpha$/Fe], where examples of $\alpha$--elements are 
O, Mg, Si, Ca, Ti, provide powerful constraints on how much of the 
Galactic halo could have been formed in dSph-sized fragments because halo 
stars have [$\alpha$/Fe] that is independent of metallicity and 
approximately equal to the theoretical average yield of type II supernovae 
(SNe).  (For reviews see Wheeler et al.~1989, McWilliam 1997).
Type II SNe explode very quickly $\sim 10^7$ yrs after stars 
form, while type Ia SNe explode afterward from $10^8$ yr to many Gyr 
after stars form. Thus the Galactic halo has been inferred to have 
formed quickly because only ejecta from short-lived type II SNe, 
and {\bf not} from long-lived type Ia SNe, were incorporated into 
most halo stars.

Instead of having short formation timescales, we now know that many of the 
dSphs in the Local Group have had surprisingly complex star-formation rates 
despite their small mass ($10^7$ to $10^8 {\rm M}_\odot$) 
and current lack of gas
(see Smecker-Hane \& McWilliam 1999, and Grebel 2000 for recent reviews). 
Color-magnitude diagrams show the Sgr dSph stars have a wide
range in age, $\sim 1$ to 15 Gyr, and  a wide range in metallicities 
from $-2 \leq \feh \leq  -0.7$ (Bellazzini et al. 1999, 
Layden \& Sarajedini 2000).  Such complex evolution should leave 
obvious signatures in the abundance ratios of the stars (Gilmore \& Wyse 
1991). Will Sgr dSph stars contain a mix of Type Ia and II ejecta?  Yes, 
if the dSph can recycle ejecta over long timescales as supported by the 
observed ranges in age and metallicity.  No, if the first Type II SNe disrupt 
the interstellar medium in less than $0.1$ Gyr and clear the way for 
subsequent SNe ejecta to escape in galactic winds, or if dSphs accrete fresh
gas that fuels star formation at later times. Therefore, determining the
abundance ratios in Sgr dSph stars also gives us unprecedented information 
on its evolution. The abundance ratios of the metal-poor stars tell us 
the initial mass function of the massive stars that exploded as Type II SNe, 
which is critical for estimating the energy available to power galactic winds. 
By modeling the measured abundances as a function of metallicity or age, we can
constrain the rate of enrichment from Type Ia and II SNe, star-formation rate, 
and the inflow/outflow of gas from the dSph.

We have obtained high dispersion spectroscopy of 14 red gaints stars in
the Sgr dSph and derived abundances for 20 elements. In a companion 
paper, we will present details of the target stars, observations, 
measured equivalent widths, and abundance analysis. In this paper we
report on our results for Fe, $\alpha$--elements, Al, Na, and neutron
capture elements (La, Y, Eu).  In addition, we derive ages for the stars
by comparing their inferred bolometric luminosity and effective temperature
to Padova stellar evolutionary models (Girardi et al.~2000). 

\section{Observations}

High resolution (R $\simeq$ 34,000) spectra of 14 Sgr dSph red giant stars were 
acquired with the Keck~I 10-meter telescope and HIRES (Vogt et al.~1994) 
from 1996 to 1998; the typical S/N per extracted pixel was $\approx$ 50.  
Program stars were selected to span the full range in color of red giant
branch stars in order to probe the full range of metallicities. 
Our sample was selected from the two fields imaged by Sarajedini \& 
Layden (1995, hereafter SL95) near the center of the Sgr dSph, and our stars were 
identified as members of the Sgr dSph based on radial velocities 
obtained by Ibata et al.~(1997). 

\begin{figure}
\plotone{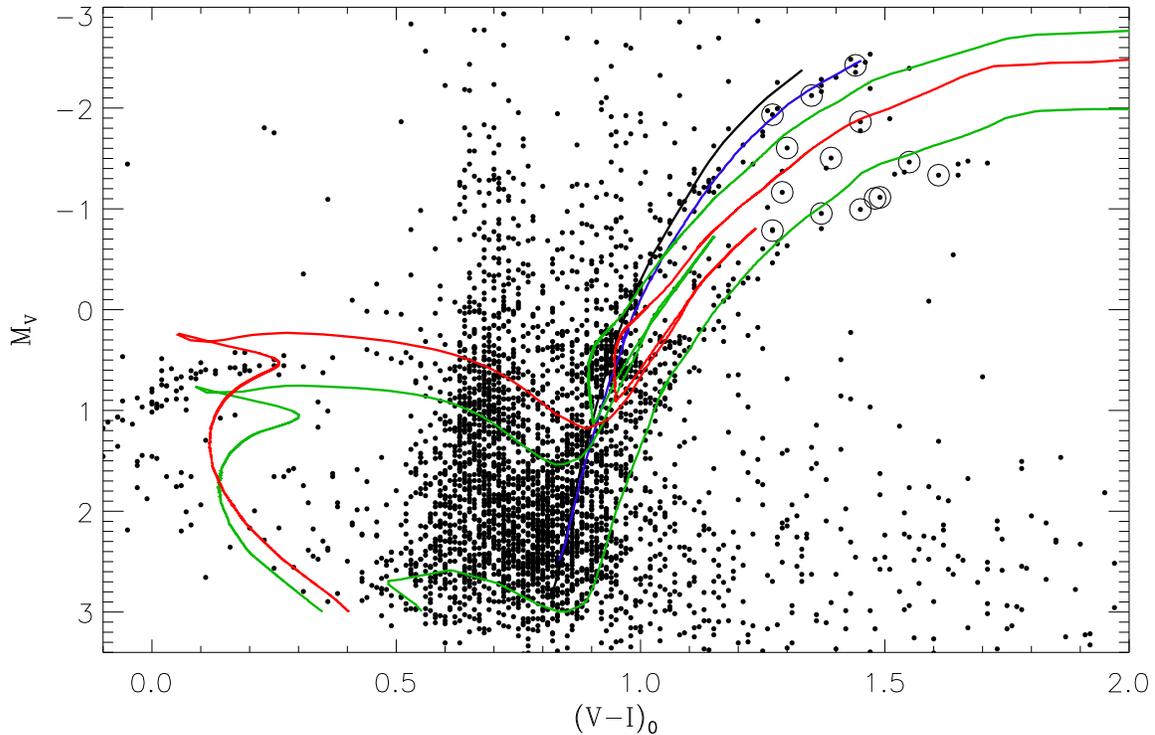}
\caption{The color-magnitude diagram for the Sgr dSph (see text for details)
with our spectroscopic targets shown as large circles. Our sample
spans the full spread in color of red giant stars in order to probe 
the full range of metallicity. Isochrones based on the Padova models 
(Girardi et al.~2000) are shown as solid lines for a variety of age 
and metallicity combinations: in black is $\feh = -1.4$ and age $= 13$ Gyr, 
in blue is $\feh = -1.0$ and age $= 5.0$ Gyr, in green are $\feh = -0.4$ 
and ages $= 1.0$ and 3.3 Gyr, and in red is $\feh = -0.1$ and age $=0.7$ Gyr. 
For the sake of clarity, evolutionary stages past the tip of the red giant 
branch are shown only for the two youngest models.}
\label{fig-VIcmd}
\end{figure}

In Figure \ref{fig-VIcmd}, we show the
dereddened color-magnitude diagram of the Sgr dSph based on the 
SL95 photometry, where we have assumed a reddening
of $E(V-I)=0.18$ and distance modulus of $(m-M)_V = 17.65$ as
derived by SL95.  The 14 stars in our spectroscopic
sample are marked by large circles.  
In order to show a fairer
representation of the stellar population in the Sgr dSph, we have 
excluded stars that are within 2 arcmin of the center of M54, which is a
globular cluster that is a member of the Sgr dSph. 
Readers should not confuse the sequence of stars located at
$((V-I)_0, M_V) \approx (0.1, 0.75)$ with a strong young population;
this sequence is the extended blue horizontal branch of M54.  However,
more extensive photometry by Bellazzini et al.~(1999) and Layden \& Sarajedini
(2000) in which they statistically subtract out the Galactic foreground
shows that a young, metal-rich, population does exist (for example,
see Bellazzini et al.'s Figure 11).  Note that the color-magnitude diagram is 
heavily contaminated by Galactic foreground dwarfs because the
Sgr fields lie at low Galactic latitude, $b = -14^\circ$. The main sequence
turnoff of the Galactic disk appears as a prominent vertical feature
at $(V-I)_0 \approx 0.7$, and lower mass main-sequence dwarfs become
more prominent in the diagram at fainter magnitudes and redder colors. 
For
\newpage
\noindent reference, we have overplotted isochrones based on the newest 
Padova models (Girardi et al.~2000) for a variety of ages and metallicities.

\section{Data Reduction and Chemical Abundance Analysis}

Extraction and wavelength calibration of the spectra was performed with 
standard IRAF routines.
Identification and measurement of the equivalent width (EW) of atomic
lines was accomplished by use of the semi-automated routine GETJOB 
(McWilliam et al.~1995).  Abundance analysis was performed using the
synthesis program MOOG (Sneden 1973) and the 64-layer Kurucz (1993) model
atmospheres.  Model atmosphere parameters for the program stars (bolometric
luminosity, $M_{\rm bol}$, effective temperature, $T_{\rm eff}$, and
gravity, $\log g$) were derived from a combination of photometric and 
spectroscopic methods; complete details will be provided in our forthcoming paper.
Abundances were derived by matching the observed line EW with
synthesis predictions; appropriate hyperfine components were employed for
affected lines when hfs constants were available. We adopt the solar abundances
as given by Grevesse \& Sauval (1998); $\epsilon ({\rm Fe}) = 7.50$.
As typically the case, systematic errors rather than random errors are
the dominant source of error in our abundances. 
The typical 1--$\sigma$ errors in the adopted stellar parameters are 
$\sigma_{T_{\rm eff}} = 70^\circ$ K, $\sigma_{M_{\rm bol}} = 0.15$ mag, 
and $\sigma_{\log  g} = 0.15$ dex.  This corresponds to typical 
1--$\sigma$ errors of $\sigma_\feh = 0.07$ dex, and 
$\sigma_\alphafe = 0.07$ dex. Typical errors for other 
element ratios are show on the following plots.

\section{Derivation of Ages}

Ages were derived for each star by comparing the star's bolometric luminosity
and effective temperature with isochrones interpolated for the star's
derived metallicity using the Padova models (Girardi et al.~2000). 
We chose to do the interpolation of the age in the
$M_{\rm bol} - T_{\rm eff}$ plane in order to avoid additional uncertainties 
introduced by adopting a specific color-temperature transformation. 
Padova models from Girardi et al.~(2000) were used to 
generate isochrones at given metallicities; Andrew Cole kindly provided
these to us.  The Padova models were calculated assuming a scaled solar 
element mix. Because our Sgr red giants show significant departures from 
solar element ratios, as we show below, we must take this into account when
deriving their ages. The $\alpha$ elements and Fe are the main sources of
opacity. Salaris et al.~(1993) show that an isochrone for
a model of a given [Fe/H] and [$\alpha$/Fe] is nearly identical to 
that of a scaled solar model with an effective metallicity equal to

\[ \feh_{\rm eff}  = \feh + \log(0.638 \, 10^{[\alpha/{\rm Fe}]} + 0.362). \]

\noindent Therefore, we interpolated the age of a star by adopting a Padova 
model with its derived $\feh_{\rm eff}$, where we 
used the average of the Si, Ca, and Ti abundances for [$\alpha$/Fe],
because these abundances are much better defined than the O abundance that
is based on the EW of only one or two lines. Errors in the derived
age of each star were calculated by propagating the typical 1--$\sigma$ errors in
$M_{\rm bol}$, $T_{\rm eff}$, [Fe/H], and [$\alpha$/Fe].

\section{Results and Discussion}

\subsection{[Fe/H] and Age Distribution} 

\begin{figure}
\plotone{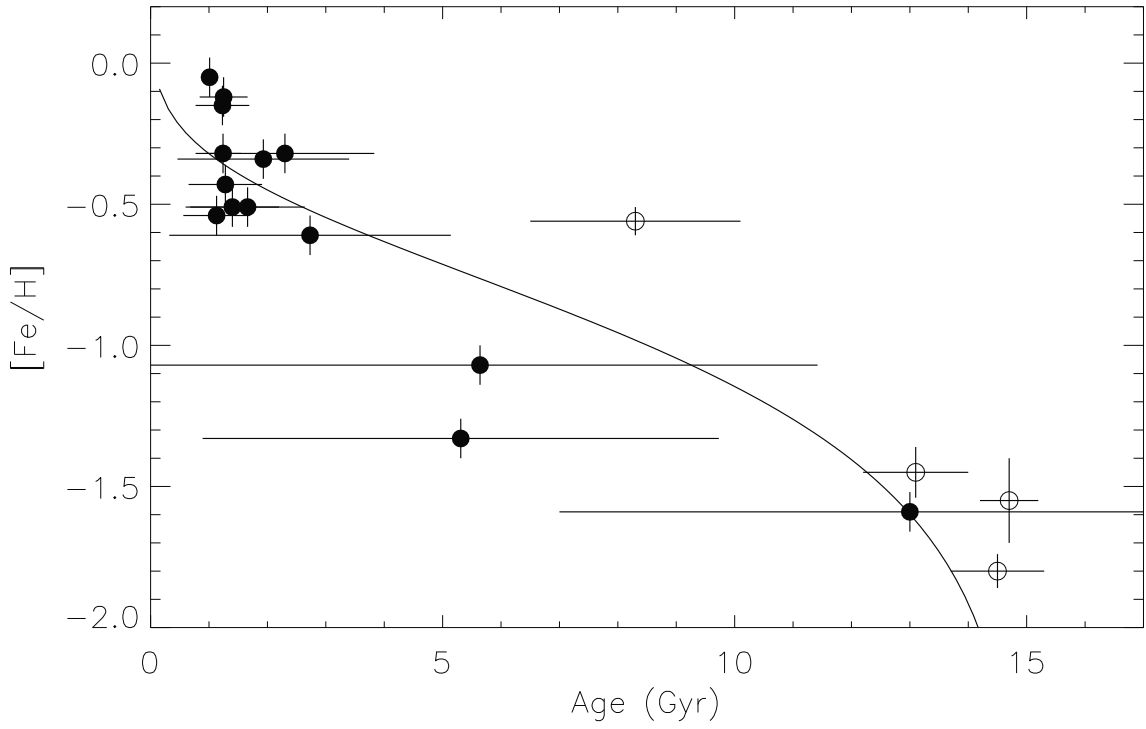}
\caption{The age--[Fe/H] relationship for the Sgr dSph. Data for our
red giant stars are shown as filled circles. Open circles show
data for 4 globular clusters that belong to the Sgr dSph (see text
for details).}
\label{fig-agefeh}
\end{figure}

Figure \ref{fig-agefeh} shows the age--[Fe/H] relationship for
our Sgr dSph stars.  The metallicities range from $-1.6 \leq \feh \leq
-0.05$.  The stars in our sample are all radial velocity members; 
furthermore, the unusual chemical compositions (discussed below) 
indicates that they are not Galactic halo, bulge, or disk interlopers.
Our high dispersion abundances confirm the existence of a solar
metallicity component as suggested by the strengths of the Ca II
near-infrared triplet lines in low dispersion spectra obtained by
Ibata and collaborators (1996, private communication to TSH).

The metallicity spread in the Sgr dSph was inferred to be large,
$-2.0 \simlt \feh \simlt -0.7$, based on the wide distribution in
colors of its red gaint stars (Bellazzini et al.~1999).
Bellazzini et al.~performed a VI-band photometric survey over a wide area
in 3 different Sgr fields. For $\sim100$ stars on the upper
red giant branch, they derived metallicities
by interpolating fiducial sequences of Galactic globular
clusters. This provided the best way of separating Sgr stars
from Galactic foreground dwarfs and the estimating the metallicity
distribution from photometry alone. They concluded that
80--90\% of Sgr stars were metal-poor, $\feh < -1$.
However our spectroscopic metallicities for most Sgr stars 
are significantly higher than this.  This discrepancy is understandable
because of the degeneracy of age and metallicity in color-magnitude diagrams.
The derived ages of our stars vary from $\sim 0.5$ to 15 Gyr, hence many
Sgr stars are significantly younger than globular clusters. 
Significantly decreasing the age of these stars would have implied a much higher
metallicity.

The age sensitivity of the red giant branch is small for ages $\simgt 3$ Gyr,
and thus our derived ages for the stars older than this have large uncertainties.
To compliment the field star data, we also show in Figure~\ref{fig-agefeh} 
data from 4 globular clusters that are members of the Sgr dSph (M54, Ter8, Arp2, Ter7).
Cluster ages were taken from Layden \& Sarajedini (2000) and are
derived from the V magnitude of the cluster subgiant branch and placed on the
age scale of the Bertelli et al.~(1994) isochrones. The Bertelli et al.~models
were the precursors of the Padova models that we use to derive the 
ages of our Sgr field stars, and their absolute age scales are roughly
similar. In order to compare with our field star metallicities derived
from high-dispersion spectra, we adopt [Fe/H] $=-1.55$ for M54 as
derived by Brown, Wallerstein \& Gonzalez (1999) from high dispersion
spectra of five M54 red giants. For the other clusters without 
high-dispersion spectroscopy, we use metallicities from
the catalog of Rutledge et al.~(1997) that are derived from 
the strengths of the Ca II infrared triplet lines in low dispersion spectra
of red giants, but calibrated to the Carretta \& Gratton [Fe/H] scale, 
which is based on high-dispersion spectroscopic abundances. 

The spread in [Fe/H] and age of Sgr dSph stars indicates that star
formation and chemical enrichment was prolonged, with a possible gap
between metal-rich and metal-poor populations.
It would be interesting to obtain high dispersion spectra of stars in
Ter 7, the cluster inferred to be $\sim 4$ Gyr younger than the other globular clusters,
as well as additional intermediate-aged and older field stars to flesh out the age-metallicity
relationship.  In Figure \ref{fig-agefeh}, the solid line shows the prediction from
a simple chemical evolution model that assumes instantaneous recycling,
a closed box, star formation has gone to completion at the present time
(i.e., the mass of gas at 15 Gyr is zero) and a star formation rate
that is constant in time. Layden \& Sarajedini (2000) suggested such a model
was a good fit to the age-metallicity relationship they derived from color-magnitude
diagram analysis. A model with a yield of  $p = 0.0033=0.17 \, {\rm Z}_\odot$ is implied
by our data although this model does not fit the data very well.  Note that this
yield is much lower than the yield inferred from stellar nucleosynthesis.
For comparison, the yield in the Galactic bulge, which is reasonably well
fit by a simple closed box model, is $p=0.7 \, {\rm Z}_\odot$ (Rich 1990)
and that implied by the the metallicity distribution function of the Galactic disk
indicates the yield is $p = 0.50 \, {\rm Z}_\odot$ (Pagel \& Patchett 1975).
Such a low yield in Sgr implies that the closed box assumption is 
probably not valid. In a simple outflow model, where the outflow rate 
is assumed to be proportional to the star formation 
rate ($\nu \psi$, where $\psi$ is the star formation rate and $\nu$ is the
dimensionless proportionality constant), the age-metallicity relationship 
is identical to that predicted by the closed box model with the 
effective yield being $p_{\rm eff} = p/\nu$.  Therefore, our Sgr results would
imply $\nu \approx 3.5$ if the true yield is the same as that in the Galaxy
and thus mass loss has been a signficant factor in the 
evolution of the Sgr dSph.

\subsection{Alpha Elements} 

In this paper, we discuss the element ratios [$\alpha$/Fe]
where $\alpha$ is the average of Si, Ca, and Ti abundances.
The O abundances are much more uncertain because they are
based on only 1 or 2 lines.

\newpage

\begin{figure}
\plotone{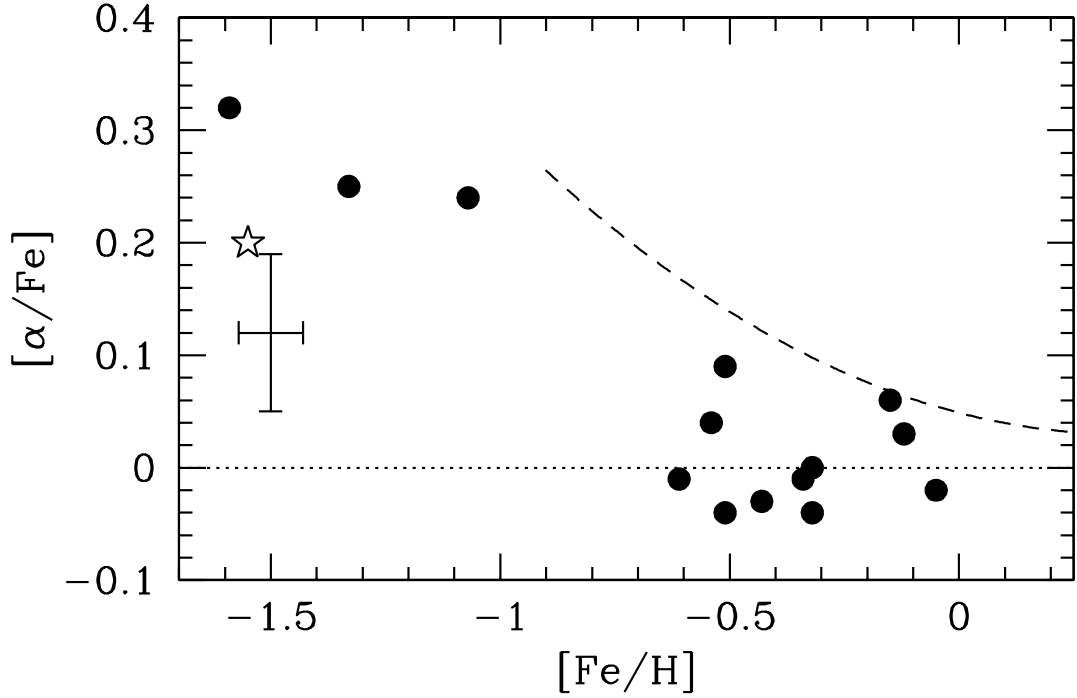}
\caption{Chemical abundances for red giants in the Sgr dSph (filled circles),
where [$\alpha$/Fe] is the average of [Si/Fe], [Ca/Fe], and [Ti/Fe].
A typical error bar shown on the lower left. The average abundance of
red giants in M54, a globular cluster in the Sgr dSph, from
Brown, Wallerstein \& Gonzalez (1999) is shown as the star symbol.
The dashed line represents the mean trend in [$\alpha$/Fe] for
stars in the solar neighborhood from Edvardsson et al.~(1993).
\label{fig-alphafe} }
\end{figure}

Figure \ref{fig-alphafe} shows that the most metal-poor Sgr dSph stars 
exhibit the enhanced [$\alpha$/Fe] ratio seen in Galactic halo, near $+0.3$ dex; 
however, above [Fe/H]$\sim$$-$1, 
it is clear that the $\alpha$-elements are deficient 
relative to the solar neighborhood at any given [Fe/H].
The standard interpretation of this observation (e.g. 
Wheeler et al.~1989, McWilliam 1997) is that the ratio of type~Ia/type~II
SNe material is larger in the Sgr dSph than the solar neighborhood above
$\feh = -1$.  This, in turn is likely due to a slower star formation 
rate, for example, a longer e-folding timescale, in the Sgr dSph.  
Such variation of $\alpha$-elements has long been 
predicted for low-mass systems (e.g. Matteucci \& Brocato 1990, 
Gilmore \& Wyse 1991).  A similar result was found for disk stars with 
large Galactocentric radii in the study of Edvardsson et al.~(1993), 
where it was concluded that the star formation proceeded at a slower pace for
the outer disk than for the inner disk.

Note that the low metallicity stars in the Sgr dSph
have [$\alpha$/Fe] ratios similar to those of Galactic halo stars,
and essentially equal to the theoretical yield of type II SNe,
means that the upper mass end of the initial mass function in the dSphs
is not significantly different than that of the Galaxy.
This is a crucial point to establish because the upper-mass end of
the initial mass function sets the amount of energy available to
power galactic winds. 

\subsection{Aluminum and Sodium}

\begin{figure}
\plotone{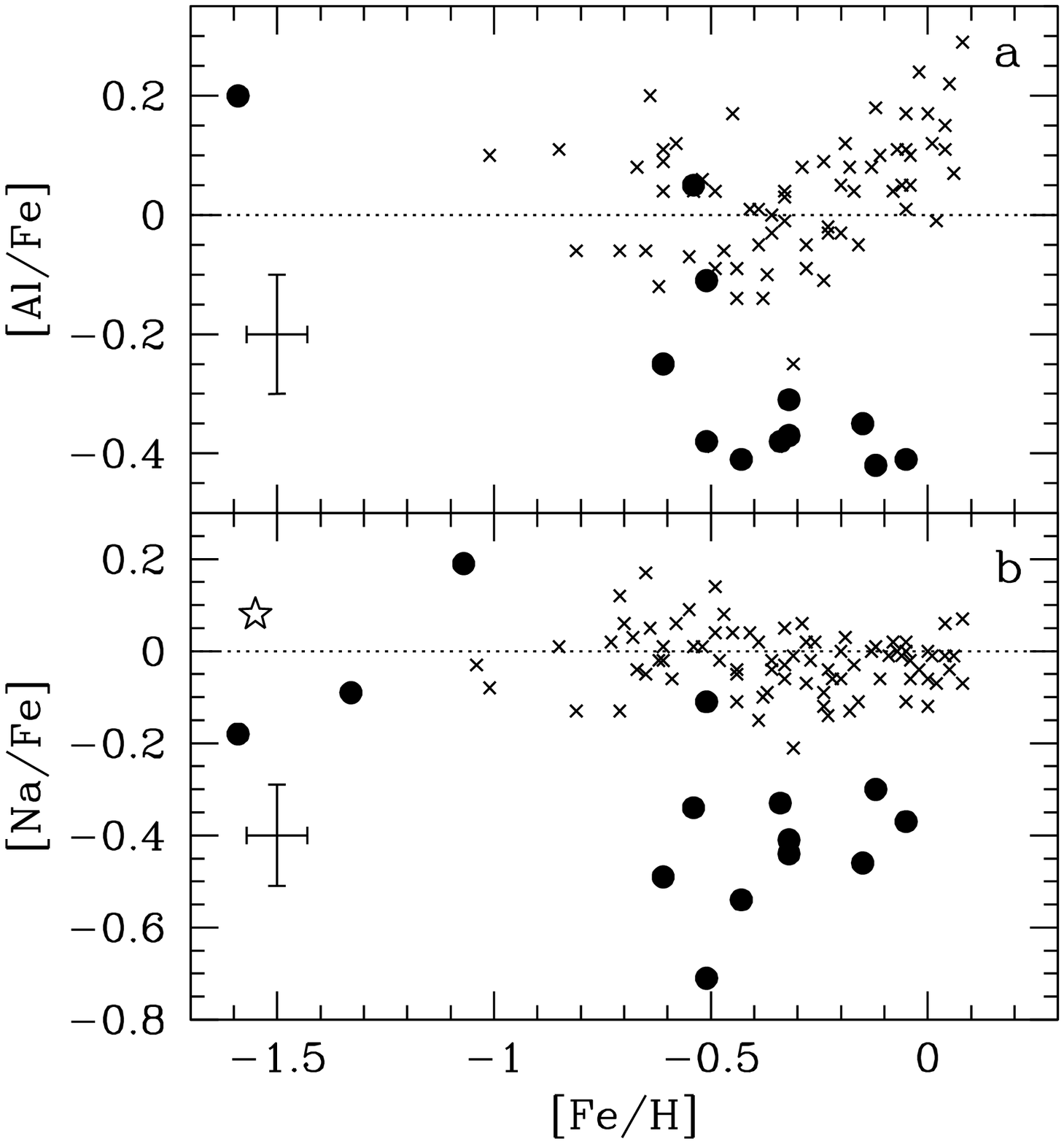}
\caption{{\bf a: }A plot of [Al/Fe] versus [Fe/H].
{\bf b: }A plot of [Na/Fe] versus [Fe/H].
Symbols are the same as in Fig.~\ref{fig-alphafe}. Crosses
represent abundances from Chen et al.~(2000) for
solar neighborhood F stars.
\label{fig-naal}}
\end{figure}

In Figure \ref{fig-naal} we show a comparison of [Al/Fe] and [Na/Fe] in the
Sgr dSph with the results for solar neighborhood stars from Chen et al.~(2000).  
The Sgr stars with [Fe/H]$<$$-$1 possess halo-like abundances
of Al and Na, [Na/Fe]$\sim$[Al/Fe]$\sim$0.  

A notable exception is star 1$-$73 with $\feh = -1.07$ that shows a 
large enhancement of Al, [Al/Fe] $ = +1.1$ dex (so large that it is 
not plotted in Figure \ref{fig-naal}), a small enhancement in Na, 
[Na/Fe] $=+0.2$, and a large deficit of O, the upper-limit being 
[O/Fe] $\simlt = -0.8$. We suggest this is due to 
proton-burning in the stellar atmosphere altering the star's primordial
abundances, depleting O and creating Na and Al, similar to the pattern
seen in some globular cluster red giants (e.g. Kraft et al.~1997).
Note that this star's derived metallicity is significantly different from
the mean metallicity of M54 stars, $\feh = -1.55$ (Brown, 
Wallerstein \& Gonzalez 1999). Thus we believe this star
is a field star and not a cluster member, although its position on the sky
and radial velocity are not significantly different from M54 stars.
If this star is a field star then the fraction of Sgr field stars showing
signs of proton burning is 7\%, although the uncertainty is $\pm 7$\%. 
This would significantly limit the amount of
material from Sgr-like dwarfs that has been incorporated into the 
Galactic halo, because no field star (with 160 surveyed to date; 
Fulbright 2000) in the Galactic halo is observed to have undergone such processing.
In the Galaxy, proton buring appears to be a phenomenon solely limited to some 
(but not all) Galactic globular cluster stars. 
Therefore additional searches for field Sgr stars with similar
signs of proton-buring would be very interesting.

For the stars with $\feh > -1$ both Al and Na are deficient relative to the
solar neighborhood trend by $\sim 0.4$ dex.  While such low [Na/Fe]
and [Al/Fe] ratios are not unheard-of in the Galaxy, they are extremely unusual,
and never found in the metallicity range, $-0.6 \simlt \feh \simlt 0$, seen 
in the Sgr dSph stars.

Apart from re-processing Ne and Mg by proton burning reactions in stellar
envelopes, Al and Na are synthesized by carbon and oxygen burning, which
only occurs in massive stars destined to end as type~II SNe.  We suggest 
that the source of the Al and Na deficiencies seen in Sgr dSph stars is due to
a paucity of nucleosynthesis products from massive stars.  This is in
qualitative agreement with the observed deficiency in [$\alpha$/Fe] ratios.
However, the Al and Na deficiencies are larger than the $\alpha$-element
lacuna, which might be understood if type~Ia SNe produce a small amount
of $\alpha$--elements as suggested, for example, by Nomoto et al.~(1984).

\subsection{Neutron-Capture Elements}

\begin{figure}
\plotone{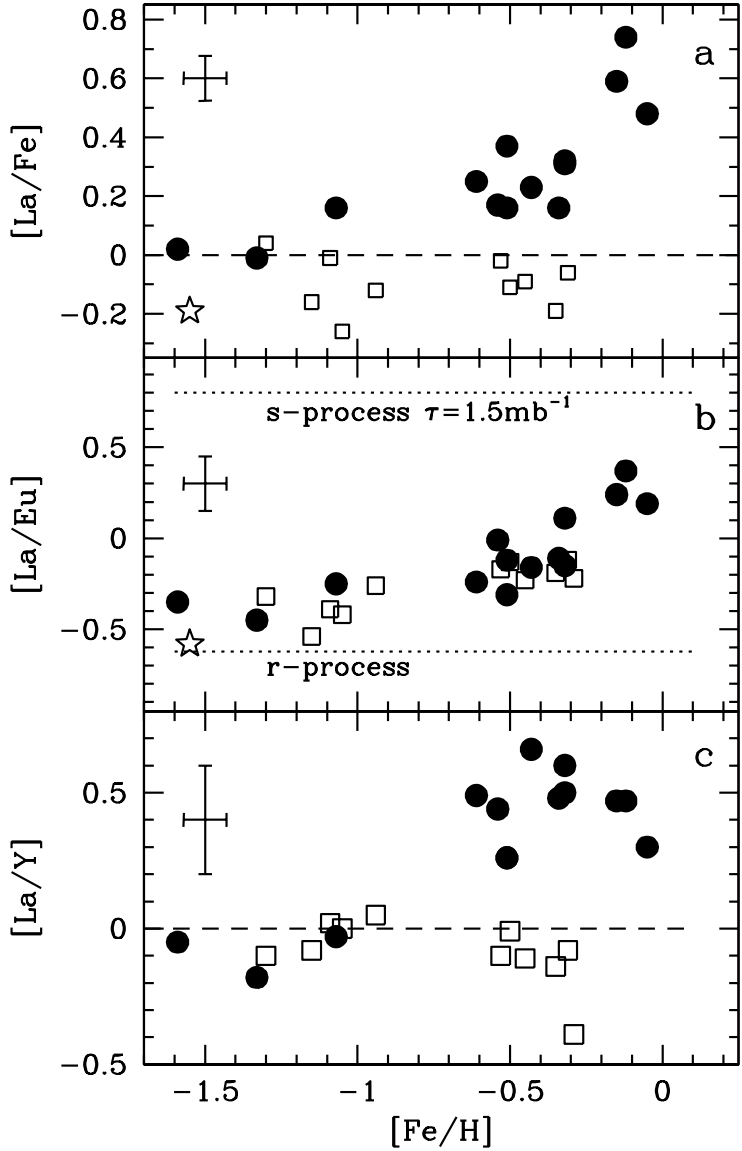}
\caption{{\bf a:} A plot of [La/Fe] versus [Fe/H].
{\bf b:} A plot of [La/Eu] versus [Fe/H].  Dotted lines indicate the
solar r-process ratio, and a pure s-process ratio from Malaney (1987).
{\bf c:} A plot of [La/Y] versus [Fe/H].
Symbols are the same as in Fig.~\ref{fig-alphafe}.
Open squares represent chemical abundances
of Galactic stars from Gratton \& Sneden (1994).
\label{fig-laeuyfe2} }
\end{figure}

In Figure~\ref{fig-laeuyfe2}a we present the trend of [La/Fe] with [Fe/H]; while the
metal-poor stars appear quite normal in this plot, the metal-rich stars
show a steady increase in La enhancement with increasing [Fe/H], up to 
[La/Fe]$\sim+0.7$ dex.  Except for very metal-poor stars, La is produced 
mostly by the s-process; the solar s-process fraction for La is estimated 
at 75\% (Burris et al.~2000).  To investigate the neutron source for La in 
the Sgr dSph stars we show the [La/Eu] ratios in Figure~\ref{fig-laeuyfe2}b; 
it is clear that [La/Eu] is increasingly dominated by the s-process at higher [Fe/H].  
In particular, the three highest [Fe/H] stars have super-solar 
[La/Eu]$\sim +0.3$ dex.  Note that normally we would like to use the 
[Ba/Eu] ratio, which is a more sensitive
discriminator of s/r-process fractions; but the Ba lines in our spectra
are so strong that they are on the flat portion of the curve of growth,
and consequently not very sensitive to abundance.
Since AGB stars are the dominant source of heavy s-process elements
at Galactic disk metallicities (see Travaglio et al.~1999), the Sgr dSph
heavy element abundance enhancements indicate a significant contribution 
from AGB nucleosynthesis, which increases with [Fe/H].

In Figure~\ref{fig-laeuyfe2}c we present a plot of [La/Y], showing that the light
s-process element yttrium is not enhanced as much as the heavy s-process
element lanthanum in the metal-rich Sgr dSph stars ([La/Y] $\sim +0.5$ dex).  
This is characteristic of metal-poor s-process environments (Busso et al.~1999).  
From Figure~16 of Busso et al.~(1999), we estimate that the
metallicity of the stars responsible for the s-process enhancement in
the metal-rich Sgr dSph population was $\feh \le -1.5$, or
at a single, higher, metallicity, near $\feh = -0.6$
(due to the bi-valued nature of the heavy/light yield function).
For AGB s-process nucleosynthesis in the metal-rich Sgr dSph stars with
$\feh \sim -0.1$ the [La/Y] ratios are expected to be $-$0.2 to $-$0.3 dex
(Busso et al.~1999), much lower than the observed value near $+$0.5 dex.
Note that the metal-poor Sgr dSph stars do not show the enhanced [La/Y]
ratios because of the importance of the r-process at $\feh \le -1$.

The observation that the s-process elements came from a significantly
more metal-poor astrophysical site than the stars themselves, rules
out the possibility that the metal-rich Sgr dSph stars polluted their own
atmospheres with s-process elements, or that s-process material was
transferred from evolved companions; it also eliminates the 
instantaneous re-cycling assumption for the chemical evolution of
the Sgr dSph.  Self pollution is also excluded because none of the 
metal-rich Sgr dSph stars are luminous enough to be on the 
thermally-pulsing AGB (TP-AGB); the log L/L$_{\odot}$
range is 2.54 to 3.07, compared to the lowest TP-AGB onset luminosity of
log L/L$_{\odot}$=3.1 from the calculations of Boothroyd \& Sackmann (1988).
Additional evidence against mass-transfer from an evolved companion comes
from the frequency of such s-process enriched stars in the Galactic disk,
at only 1 to 2\% (MacConnell \& Frye 1972); it seems very unlikely that
all the metal-rich Sgr dSph stars could have come from such a rare population.

The observation that all the metal-rich Sgr dSph stars show a similar
enhancement in the [La/Y] ratio is evidence that the progenitor AGB stars
responsible for the s-process elements were of a similar metallicity;
the simplest explanation is that the progenitors were from the
metal-poor population.  

Since AGB s-processing occurs in the approximate mass range
$ 1.2 {\rm M_\odot <  M } < 5 {\rm M_\odot}$, 
we conclude that there was an extended period of
star formation and chemical enrichment in the Sgr dSph which lasted
from $\sim$0.5 to a few Gyr.  

The metal-rich Sgr dSph stars, $\feh \simgt -0.7$, have derived
ages from 0.5 to 3 Gyr, while the metal-poor stars range from
5 to 13 Gyr; this is consistent with the expectation that ages decrease
with increasing [Fe/H], and that the metal-poor population is similar to the
Galactic halo.  The derived ages are in qualitative agreement with a period of 
extended star formation as indicated by the abundances of neutron-capture elements.

We note that similar s-process enhancements have been seen in stars 
of the globular cluster Omega Cen (e.g. Smith et al.~2000), but at lower
metallicity; those observations were attributed to AGB nucleosynthesis by 
M $< 3 {\rm M_\odot}$ stars.

Thus, the detailed abundances of neutron capture elements in the Sgr dSph
indicate that the old, metal-poor, population was the dominant source
of the heavy elements for the metal-rich population.

This is quite contrary to the situation in the solar neighborhood, where
the G-dwarf problem is the observation of the absence of a significant
metal-poor thin disk.

The unusual heavy element abundance patterns in metal-rich Sgr dSph stars
are also consistent with considerable mass loss from the galaxy
after the initial formation, as indicated by the [Fe/H] distribution; if
this were not the case, the high [La/Y] ratios from the metal-poor component
would have been overwhelmed by the products of a more populous metal-rich
population.

\subsection{Comparison with Other Studies}

Bonifacio et al.~(2000) derived chemical abundances for 2 red giants in 
the Sgr dSph from high-dispersion spectra obtained with UVES on the
ESO 8.2-meter telescope. Their derived metallicities are [Fe/H] $= -0.28$
and $-0.21$, similar to our dominant group at high metallicity.
The abundance ratios they derive fit very well onto the trends defined by
our data. They note that the unusual abundances found for the high
metallicity Sgr stars are very much like those seen in young supergiants
in the Large and Small Magellanic Cloud (c.f., Hill 1997 and references
therein). Therefore, our explanation of Sgr's unusual abundance 
variations -- namely, slow evolution of the star formation rate accompanied
by mass loss and chemical enrichment of the young, metal-rich population by the old, 
metal-poor population  -- is probably applicable to the Magellanic Clouds. 

It is curious that the youngest stars in the Sgr dSph have approximately the
same metallicities, as measured by [Fe/H], as the young stars in the
LMC despite the fact that the total absolute magnitude, M$_{\rm V}$, for the 
LMC is more than 5 magnitudes brighter than the Sgr dSph. Is this an artifact
of star formation running to completion in Sgr dSph, or is that 
much of Sgr's luminosity has already been stripped by tidal interactions 
with the Galaxy?

\subsection{The Complex Evolution of the Sgr dSph}

A detailed modeling of the age and metallicity relationship is outside
the scope of the present data. Ideally, one would want to obtain spectroscopy
for a large number of stars and simultaneously analyze the abundances and
well-populated color-magnitude diagrams. 

\begin{figure}
\plotone{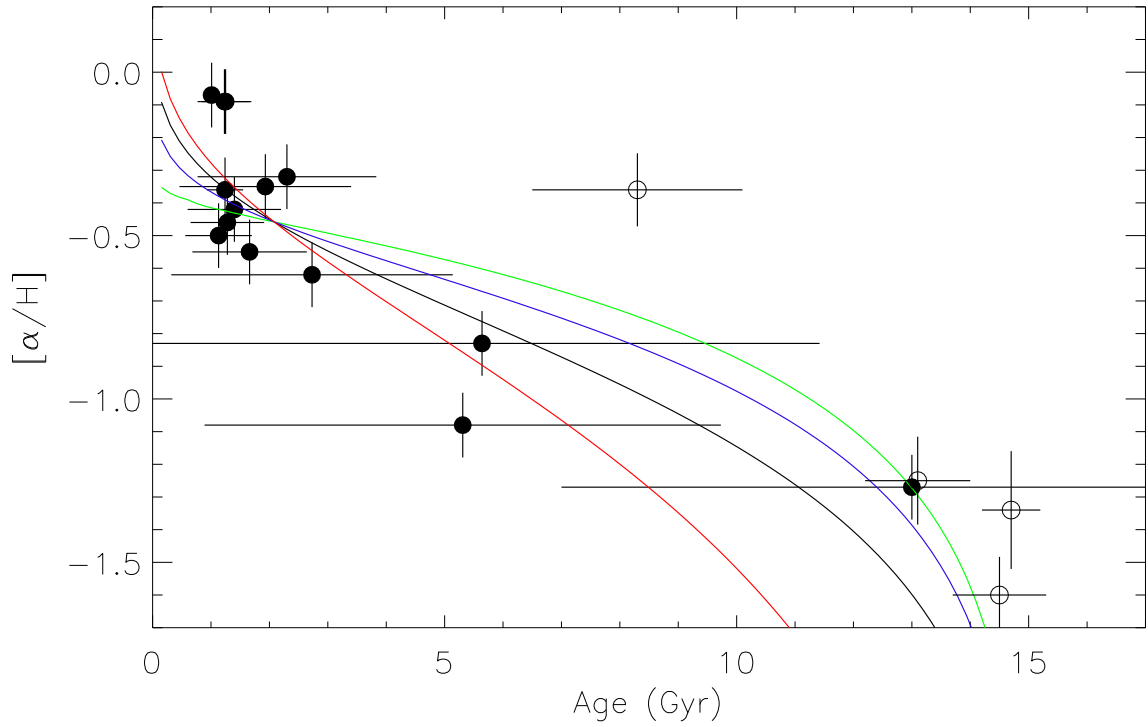}
\caption{The age--metallicity relationship for the Sgr dSph. Filled circles show
the data for our 14 red giant stars. Open circles show data for
4 globular clusters that are members of the Sgr dSph.
Simple chemical evolution models that assume instantaneous recycling
are shown as solid lines. See text for details.
\label{fig-agealphah} }
\end{figure}

However we can glean more information about the evolution of the 
Sgr dSph by comparing our data to some simple models.  In Figure \ref{fig-agealphah},
we plot the age and [$\alpha$/H] abundance, which is more likely to be a good candidate
for the instantaneous recycling approximation than [Fe/H] because of the delayed
explosion of type Ia SNe which generate significant amounts of Fe but little
$\alpha$-elements. For each of the globular clusters, we have assumed 
[$\alpha$/Fe] = 0.21, which is the value found by Brown, Wallerstein \& 
Gonzalez (1999) for M54.

Figure \ref{fig-agealphah} illustrates the results of simple outflow chemical 
evolution model with instantaneous recycling. For comparison, they have all been 
normalized to pass through the same point. Each model assumes that star formation
goes to completion at the present time of $t=15$ Gyr, but adopts different 
prescriptions for the star formation rate, $\psi(t)$. 
The black line shows the prediction assuming $\psi$ is constant in time. Results
for an exponentially decreasing star formation rate of the form
$\psi(t) \propto e^{-t/\tau}$ are illustrated by the green line for $\tau = 5$ Gyr
and the red line for $\tau = 1$ Gyr. A model with an increasing star formation rate
of the form $\psi \propto t$ is illustrated by the red line.
Note that none of these simple outflow models accurately reproduce the 
observed age-metallicity relationship, but the sharp increase in [$\alpha$/H] for
ages $\simlt 5$ Gyr would imply an increasing, as opposed to decreasing, star
formation rate. The most plausible model might well involve a discontinuous
star formation rate. Episodic star formation events have been found for a
number of dSphs in the Local Group with the most stunning example being
the Carina dSph (c.f., Hurley-Keller et al.~1998, Smecker-Hane \& McWilliam 1999).
Further work on the Sgr dSph will benefit greatly from new generations of
spectrographs, such as the MIKE fiber-fed echelle spectrograph being developed
for the 6.5-meter Magellan Telescopes, which will allow one to 
obtain high-dispersion spectra for $\sim 100$ stars simultaneously. 
Modeling of the color-magnitude diagrams, [Fe/H], and chemical abundance ratios 
promise to yield powerful constraints on the physical processes that
regulated the complex evolution of dSphs.
 
\section{Summary and Conclusions}

Our observations of the Sgr dSph are consistent with prolonged chemical
enrichment with significant mass loss.  The radial velocities and chemical
composition firmly establish our sample as bona fide Sgr dSph members with
a metallicity range of $-1.6 \leq \feh \leq -0.05$ dex and
an corresponding age range of $\sim$13 to 0.5 Gyr.

While the composition of the metal-poor, $\feh < -1$, Sgr dSph stars 
resembles the Galactic halo stars, the metal-rich component shows
a very unusual composition.  The [$\alpha$/Fe] ratios are, on average,
slightly sub-solar in the metal-rich sub-sample ($\feh \sim -0.3$ dex),
which implies a low type~II/type~I SN ratio, and can most reasonably be
understood as due to a relative paucity of type~II SNe, consistent with
expectations for a slow, or possibly episodic, star-formation rate.

In the metal-rich Sgr dSph stars the abundances of Al and Na are 
both deficient by $\sim$0.4 dex, relative to iron, which has not previously
been seen at this [Fe/H].  Since it is necessary for a star
to go beyond carbon burning to produce large amounts of Al and Na (ignoring
the re-distribution of Mg and Ne by proton burning), we interpret the observed
Al and Na abundance deficiencies as a consequence of the low frequency of
type~II SNe.

Enhancements of neutron-capture heavy elements, which increase with [Fe/H],
and the s-process signature seen in the [La/Eu] ratios,
indicate the importance of AGB nucleosynthesis in the metal-rich Sgr dSph
population.  However, the [La/Y] ratios suggest a nucleosynthetic origin
from sites of much lower metallicity than the stars themselves.  These
observations may be understood if heavy elements in the metal-rich population
were dominated by the low-mass, long-lived, AGB stars of the old, metal-poor
population.  The relatively long-lived AGB progenitors require that 
star formation in Sgr dSph took place over an extended period of time,
at least 0.5 Gyr, and probably a few Gyr.  This assertion is supported
by age estimates for our sample of Sgr dSph stars, based on Padova isochrones,
which span the range from 0.5 to 13 Gyr.

\acknowledgments

We gratefully acknowledge financial support from the NSF through grants 
AST-9619460 and AST-0070895 to TSH, and AST-9618623 and AST-0098612 to
AM.  We wish to extend special thanks to the staff at Keck Observatory
for their excellent observing support and to the people of Hawaiian ancestry 
on whose sacred mountain we were privileged to be guests.

\end{document}